# Astrometrically registered maps of $H_2O$ and SiO masers toward VX Sagittarii

Dong-Hwan Yoon[1,2], Se-Hyung Cho[2,3], Youngjoo Yun[2], Yoon Kyung Choi[2], Richard Dodson[4], María Rioja[4,5], Jaeheon Kim[6], Hiroshi Imai[7], Dongjin Kim[3], Haneul Yang[1,2] & Do-Young Byun[2]

The supergiant VX Sagittarii is a strong emitter of both $H_2O$ and SiO masers. However, previous VLBI observations have been performed separately, which makes it difficult to spatially trace the outward transfer of the material consecutively. Here we present the astrometrically registered, simultaneous maps of 22.2 GHz $H_2O$ and 43.1/42.8/86.2/129.3 GHz SiO masers toward VX Sagittarii. The $H_2O$ masers detected above the dust-forming layers have an asymmetric distribution. The multi-transition SiO masers are nearly circular ring, suggesting spherically symmetric wind within a few stellar radii. These results provide the clear evidence that the asymmetry in the outflow is enhanced after the smaller molecular gas clump transform into the inhomogeneous dust layers. The 129.3 GHz maser arises from the outermost region compared to that of 43.1/42.8/86.2 GHz SiO masers. The ring size of the 129.3 GHz maser is maximized around the optical maximum, suggesting that radiative pumping is dominant.

[1] Astronomy Program, Department of Physics and Astronomy, Seoul National University, 1 Gwanak-ro, Gwanak-gu, Seoul 08826, Korea. [2] Korea Astronomy and Space Science Institute, 776 Daedeokdae-ro, Yuseong-gu, Daejeon 34055, Korea. [3] Department of Astronomy, Yonsei University, 50 Yonsei-ro, Seodaemun-gu, Seoul 03722, Korea. [4] International Center for Radio Astronomy Research, M468, The University of Western Australia, 35 Stirling Highway, Crawley, WA 6009, Australia. [5] OAN (IGN), Alfonso XII, 3 y 5, 28014 Madrid, Spain. [6] Shanghai Astronomical Observatory, Chinese Academy of Sciences, 200030 Shanghai, China. [7] Institute for Comprehensive Education, Kagoshima University, Korimoto 1-21-30, Kagoshima 890-0065, Japan. Correspondence and requests for materials should be addressed to S.-H.C. (email: cho@kasi.re.kr)

VX Sagittarii (VX Sgr) is a red supergiant with a semi-regular variable period of 732 days[1]. The distance measured from SiO maser proper motion is 1.57 kpc[2]. The photospheric diameter is 8.82 mas[3] (13.85 AU) at 2.0 μm. This star shows a heavy mass loss of about $2.5 \times 10^{-4} M_\odot$ year$^{-1}$ [4]. It is well known that VX Sgr hosts strong OH, $H_2O$, and SiO maser emitters, compact enough for very long baseline interferometry (VLBI) observations[5–7]. The SiO masers are located at 2–4 stellar radii from the stellar surface, inside the dust-formation layer, while the 22.2 GHz $H_2O$ maser is located outside the dust layer[8,9], which is undergoing a radial acceleration. With the precise astrometrical registration of the SiO and $H_2O$ masers that are observed simultaneously, we can directly compare the properties of these masers on the scales of the individual maser gas clumps for tracing the mass transfer between these layers.

The Korean VLBI Network (KVN) employing a unique quasi-optics for simultaneous observations of K (21.3–23.3), Q (42.1–44.1), W (85–95), and D (125–142 GHz) bands[10] enables us to perform the combined studies of $H_2O$ and SiO masers toward VX Sgr. This paper presents the result of the astrometrically registered, simultaneous maps of the 22.2 GHz $H_2O$ and 43.1/42.8/86.2/129.3 GHz SiO masers using the non-integer source frequency phase-referencing (SFPR) method[11–13]. This results provide observational evidence for a break in spherical symmetry between the SiO and $H_2O$ maser zone. The 129.3 GHz SiO maser from VX Sgr shows that radiative pumping is dominant, arising from the outermost region compared to the 43.1/42.8/86.2 GHz SiO masers.

## Results

**$H_2O$ and SiO maser maps and their morphological differences**. Figure 1 shows the astrometrically registered integrated intensity map of the $H_2O$ and SiO maser lines observed on March 27, 2016 (optical phase $\varphi = 0.67$) toward VX Sgr[14]. The distributions of the SiO masers show a typical ring-like structure, while that of the $H_2O$ maser shows an asymmetric structure spread slightly in the NW and SE direction in ~350 mas and relatively dense distributions in the NE and SW direction in ~270 mas. These results are consistent with those of multi-element radio linked interferometer network (MERLIN) and very long baseline array (VLBA) observations[15,16]. In addition, there are few $H_2O$ maser features in the southern direction in contrast to those of the SiO masers.

Figure 2 shows the position–velocity spot maps according to each maser lines. The $H_2O$ maser features show a wide-angle NE–SW biconical structure extended in slightly the NW and SE direction with respect to the position of central star (the mark "x"). The blue-shifted maser spots are dominant in the eastern part, and the red-shifted maser spots are dominant in the western part. In the case of the SiO maser, the maser components we see are mainly pumped in the tangential direction, so they show a ring-like structure in the line-of-sight. However, their partially clumped structure rather than a fully populated ring and various velocity ranges show an incoherency of the maser spot distribution. The pulsation of the stellar photosphere propagates through the SiO maser region, where dust starts to form. $Al_2O_3$ is likely to be the first nuclei created, at least in some lower-mass stars, and the associated SiO maser monitoring has been used to relate their appearance to the kinematics within a few $R_*$[17]. Each SiO maser line is distributed irregularly at any single epoch and, over the pulsation cycle, can show outflow and inward motion in different regions[18], possibly associated with inhomogeneous dust formation. This would lead to local differences in the efficiency of radial acceleration through radiation pressure on grains and could be the cause of the irregular appearance of the 22.2 GHz $H_2O$ masers above the dust-formation layers. Residual stellar pulsations may also affect the inner, collisionally pumped $H_2O$ masers. The combination of our various SiO observations provides an almost perfect ring, suggesting that the average stellar wind is spherically symmetric, but the 22.2 GHz $H_2O$ maser have an NE–SW axis of the dipole magnetic field, possibly magnetically influenced[16].

**Different locations among SiO masers**. We can directly compare the maser spot distributions among the 43.1/42.8/86.2/129.3 GHz SiO masers (Table 1) in the astrometrically registered maps of Fig. 3. We can confirm the previous characteristics of the maser features and distributions according to each maser transition. In order to determine the position of the central star, we performed the ring fitting based on all spot distributions of the 43.1/42.8/86.2/129.3 GHz SiO masers. We used all of the SiO maser spots to improve the reliability and position accuracy. The mark "x" indicates the position of the central star. The ring radius obtained with the ring fitting in Fig. 2 is compared with that of the Gaussian fitting for the spot distribution histogram. Their radii obtained from both methods are consistent within the errors. Interestingly, the 86.2/129.3 GHz maser spots are located at outer regions from the central star than the 43.1/42.8 GHz maser spots (Fig. 3, Table 1). This trend has persisted at other epochs as shown in Fig. 4[14]. Therefore, we can confirm that the 42.8 GHz SiO maser is located inside the 43.1 GHz maser, the 86.2 GHz maser is located at the outer region of the 43.1 GHz maser, and the 129.3 GHz maser is located at the outermost region toward VX Sgr.

**Observational evidences of SiO maser pumping mechanism**. The different maser spot distributions among the 43.1/42.8/86.2/129.3 GHz SiO masers provide important information on the maser pumping mechanism. It is well known that the vibrational state $v = 1$ (43.1 GHz) and $v = 2$ (42.8 GHz) SiO masers at the same rotational transition $J = 1–0$ show similar distributions with overlapping features and the $v = 2$ SiO maser arises from the somewhat inner region of the $v = 1$ SiO maser in many previous

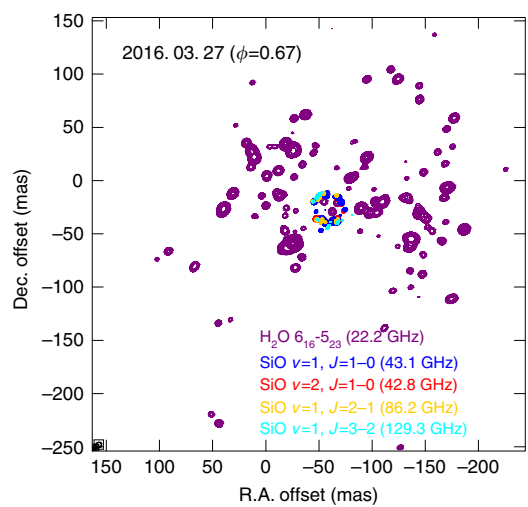

**Fig. 1** Astrometrically registered velocity-integrated intensity maps of 22.2 GHz $H_2O$ and 43.1/42.8/86.2/129.3 GHz SiO masers obtained from the SFPR technique toward VX Sgr (in press manuscript[14]). The rms noise levels on the maps are 115.4, 18.01, 15.45, 33.69, and 11.59 Jy beam$^{-1}$ m s$^{-1}$, respectively, in order of the maser transition. The peak flux values are 573.62, 55.65, 38.08, 61.70, and 1.85 Jy beam$^{-1}$ km s$^{-1}$. The contour levels are adopted at 10 12 15 20 25 30 50... for multiples of each rms value

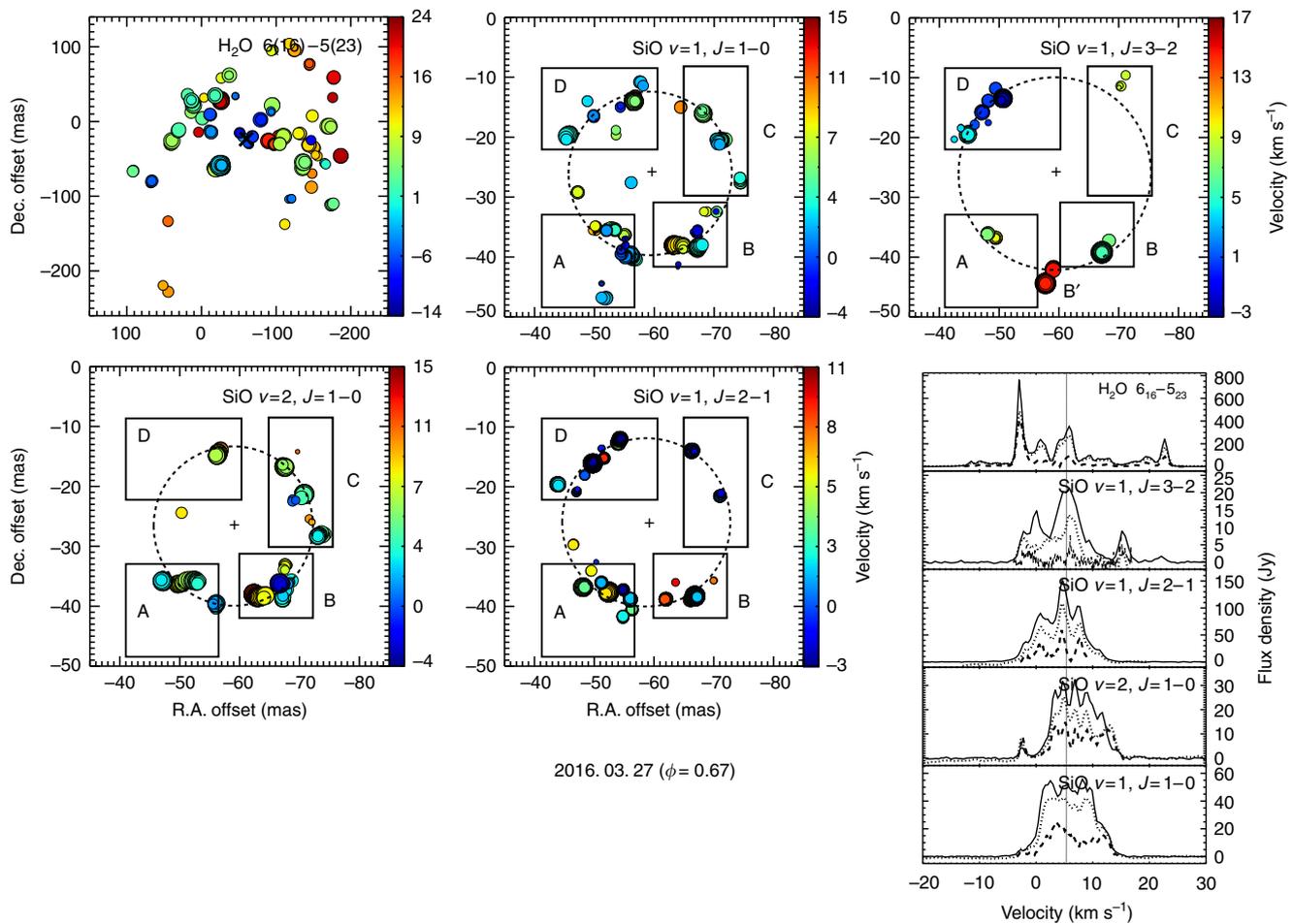

**Fig. 2** The position−velocity spot maps and flux spectra. The color of the spot indicates the local-standard-of-rest velocity and the size of spot the logarithmic scale of the intensity. The dashed circle is the ring fitted to the SiO maser spot distribution. The center of each ring is marked with "+". The position of the central star is marked with "x" on the H$_2$O distribution and assumed to be the ring center of all four SiO maser lines combined in Fig. 3. The spectra represent the single dish (solid, April 5, 2016), total power (dotted), and recovered flux (dashed) in VLBI observations (Method). The gray vertical line represents the stellar velocity (5.3 km s$^{-1}$)[5]

| Table 1 SiO maser ring-fitting results around VX Sgr on March 27, 2016 ($\varphi = 0.67$) | | | | | | | |
|---|---|---|---|---|---|---|---|
| SiO transition | R.A. offset[a] | Dec. offset[a] | Ring radius[b] | | Gaussian fit | Converted coordinate (J2000)[c] | |
| | (mas) | (mas) | (mas) | (AU) | (mas) | R.A. | Dec. |
| $v=1, J=1-0$ (43.1 GHz) | −59.281 | −26.069 | 13.63 ± 0.37 | 21.40 ± 0.58 | 13.14 | 18:08:04.0457308 | −22:13:26.626069 |
| $v=2, J=1-0$ (42.8 GHz) | −58.898 | −26.675 | 13.31 ± 0.73 | 20.90 ± 1.15 | 12.90 | 18:08:04.0457584 | −22:13:26.626675 |
| $v=1, J=2-1$ (86.2 GHz) | −58.940 | −26.036 | 14.10 ± 0.11 | 22.14 ± 0.17 | 14.08 | 18:08:04.0457553 | −22:13:26.626036 |
| $v=1, J=3-2$ (129.3 GHz) | −59.384 | −26.013 | 16.08 ± 0.81 | 25.25 ± 1.27 | 15.52 | 18:08:04.0457234 | −22:13:26.626013 |
| Four SiO masers | −58.720 | −26.104 | 14.25 ± 0.58 | 22.37 ± 0.91 | — | 18:08:04.0457712 | −22:13:26.626104 |

[a]The offset value represents the position difference of the central star with respect to the observed Hipparcos coordinates (Methods)
[b]The astronomical unit (AU) was calculated using the distance of 1.57 kpc for VX Sgr[2]
[c]The coordinates are converted from the observed Hipparcos coordinates based on the R.A and Dec. offsets

VLBI observations[19–21], which is consistent with the theoretical excitation conditions of the $v=2$ maser[22].

However, the SiO masers of the higher rotational transitions at the same vibrational state $v=1$ (SiO $v=1$, $J=1-0$, $J=2-1$, $J=3-2$) shown in Fig. 3 and Table 1 arise from the outer region, i.e., the higher rotational transition masers arise further from the central star. In the case of the SiO $v=1$, $J=1-0$ (43.1 GHz) and $J=2-1$ (86.2 GHz) masers, shock-enhanced simulation models[23] accurately predict the larger radii of the 86.2 GHz SiO maser compared with those of the 43.1 GHz SiO maser as a function of the stellar pulsation phase. Their model is also supported by our observational results in which the 86.2 GHz SiO maser intensity is stronger than that of the 43.1, 42.8 GHz SiO masers in VX Sgr. The predominantly collisional pumping model predicted that the 86.2 GHz SiO maser amplification is greater than that of the 43.1 GHz SiO masers for intermediate densities[24]. As an another possibility, the line overlap[25] can be proposed for explaining the larger ring size of 86.2 GHz maser compared to that of 43.1 GHz

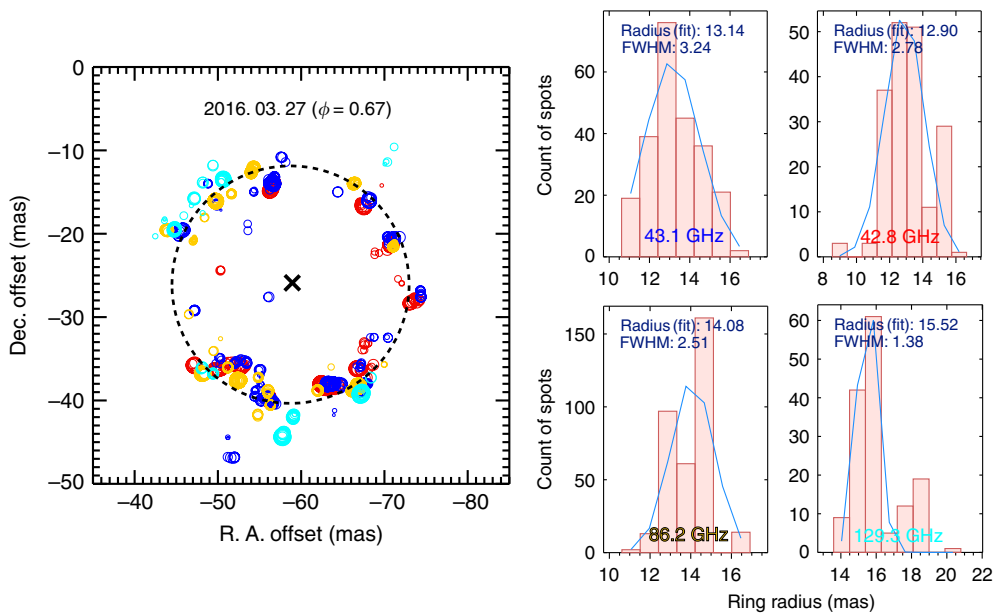

**Fig. 3** Registered maps of the SiO maser spots and the histograms of their radial distributions. The blue, red, yellow, and cyan colors indicate 43.1, 42.8, 86.2, and 129.3 GHz SiO masers, respectively. The dotted circle shows the result of ring fitting using all the maser spot distributions of the four maser lines. The mark "x" indicates the ring fitting center assumed to be the position of the central star. Histograms represent the radial distribution of the maser spots with respect to the central star. The ring radius and FWHM derived from the Gaussian fitting are represented in each panel

maser in oxygen-rich star VX Sgr. However, radiative pumping model including the line overlap needs to reproduce the strong intensity of the 86.2 GHz SiO maser compared to that of the 43.1 GHz maser.

The 129.3 GHz SiO maser is clearly imaged and located at the outermost region with the largest ring radius compared to that of the 43.1/42.8/86.2 GHz SiO maser lines (Figs. 3 and 4). The single-dish spectrum (Fig. 2) of the 129.3 GHz maser shows more complex features and a wider velocity range (form −5 to 18 km s$^{-1}$). The variation in line profile and peak velocity as a function of the stellar phase in the 129.3 GHz maser was different from that of the 86.2 and 43.1 GHz masers in single-dish monitoring observations[26]. These results may originate from the difference in their excitation conditions as discussed below.

**Pumping mechanism for the 129.3 GHz SiO maser.** We assumed that the masing regions of higher rotational transitions would be closely related to the optical depth of the circumstellar envelopes. The SiO maser population inversion (whether radiatively or collisionally pumped) is caused by self-trapping of photons corresponding to the $\Delta v = 1$ ro-vibrational transition[27]. At large opacities, spontaneous de-excitation of the SiO molecule via the $\Delta v = 1$ ro-vibrational transition becomes more difficult at higher rotational transitions[25]. This is possible even for radiative pumping in flattened regions such as in thin shells, which can be optically thick tangentially and thin radially. Thus population inversion and amplification of the higher rotational transition SiO masers is favored at progressively higher radii. Our results show a greater increase in radii and other differences between $J = 3-2$ and $J = 2-1$ masers, compared with those between $J = 2-1$ and $J = 1-0$ (Figs. 3 and 4). Thus the higher rotational transition masers in the SiO $v = 1$, $J = 1-0$, $J = 2-1$, $J = 3-2$ maser lines seem to arise further from the central star. However, we cannot exclude the radiative models including line overlap[25] for the further distance of the SiO $v = 1$, $J = 3-2$ maser from the central star. In addition, the ring radius of 129.3 GHz maser increases

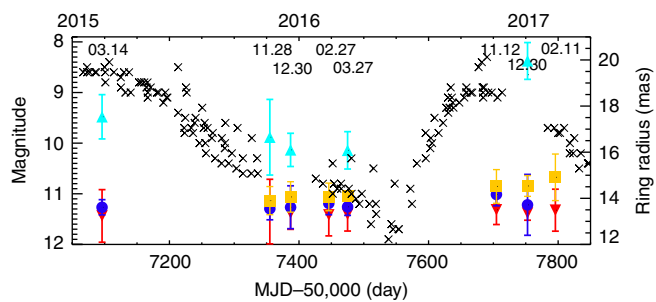

**Fig. 4** Variation of the SiO ring radius with the optical light curve (in press manuscript[14]). The gray cross indicates the optical light curve from American Association of Variable Star Observers (AAVSO; https://www.aavso.org). Color symbols indicate the SiO maser ring radius according to the different transitions and epochs. Cyan triangle, yellow square, blue circle, and red inverted triangle are in the order of 129.3, 86.2, 43.1, and 42.8 GHz SiO maser lines (Table 1, Supplementary Tables 2 and 4)

near optical maximum as shown in Fig. 4, which displays the ring radius variation of the 129.3 GHz maser at multiple epochs. This fact directly supports the radiative pumping for the 129.3 GHz maser and also suggests that the radiative pumping is more dominant than the collisional pumping in the higher $J = 3-2$ rotational transition for VX Sgr differently from the $J = 2-1$ maser.

**Discussion**

In the viewpoint of the collisional pumping of SiO masers caused by shocks, one can expect to trace the outward shock propagation near the photosphere[28] to the dust-forming layers based on the fact that the higher rotational transition masers arise further from the central star. As a future work, monitoring observations of the 129.3 GHz SiO maser together with 43.1 and 86.2 GHz SiO masers will allow us to trace the stratification structure of the excitation conditions, and inward/outward motions of maser clumps in different regions possibly associated with local shock and inhomogeneous dust formation.

Finally, our simultaneously astrometrically registered maps of the SiO and $H_2O$ masers will provide important observational constraints for the local difference in the efficiency of radial acceleration through radiation pressure on grains and the cause of the irregular appearance of the 22.2 GHz $H_2O$ masers above the dust-formation layers. In addition, the variation of the 129.3 GHz SiO ring radius with the optical light curve suggests that radiative pumping is dominant in the red supergiant VX Sgr.

## Methods

**KVN observations**. Simultaneous VLBI monitoring observations of $H_2O$ $6_{16}$–$5_{23}$ (22.235080 GHz), SiO $v = 1, 2, J = 1$–$0$, SiO $v = 1, J = 2$–$1, 3$–$2$ (43.122080 GHz, 42.820587 GHz, 86.243442 GHz, and 129.363359 GHz) maser lines were performed toward VX Sgr, one of the KVN key science project sources[29]. The observations were carried out about every 2 months from 2014 November to 2017 May. Here we report the observations on March 27, 2016 ($\varphi = 0.67$), which show the successful 4-band non-integer SFPR maps and those of 4 epochs, which show the single-band map of 129.3 GHz SiO maser (March 14, 2015, November 28, 2015, December 30, 2015 and December 30, 2016). The observed coordinate of VX Sgr is R.A = 18:08:04.05, Dec. = −22:13:26.6 from Hipparcos main catalog[30].

The passed signal is recorded onto the 1 Gbps Mark 5B (MK5B) recorder with 16 base band channels (BBCs) and 16 MHz bandwidth (512 channels) for each. We set the 6 intermediate frequencies of BBCs in the K and Q band, and 2 in the W and D band and the frequencies are arranged randomly in order to avoid high side peaks. Recorded data were correlated by the Distributed FX[31,32] software correlator. The synthesized beam size, position angle, and system noise temperature obtained from the observation for each frequency are listed in Supplementary Table 1.

We used fringe finder and continuum delay calibrator sources J1743–0350 and J1833–2103, respectively. Observations were performed over 5 h with 2 min scans for the target to continuum delay calibrator. Fringe finder calibrator source was observed every 1 h to exclude an instrumental delay and bandpass calibration. Amplitude calibration was done from the Astronomical Image Processing System (AIPS) task APCAL that measured the system noise temperature and gain variation data. There are no bright continuum delay calibrators stronger than 500 mJy (recommend for observations using 1 Gbps recording mode in KVN) and a separation angle within 4° from VX Sgr. Our delay calibrator J1833–2103 is strong but it has a complicated structure of a gravitational lensing[33] and 6.06° separation angle from the target source.

**SFPR data reduction**. The data reduction was performed using the AIPS package. Basic data treatment followed the standard procedure for the phase-referencing line imaging method in the K band[13]. First, we used the bright fringe finder to remove the residuals of large group delays and delay rates in the target itself. Second step is to trace the residuals of fringe phases using a nearby continuum delay calibrator, whose solutions were applied to the visibilities of the target. This step enables us to compensate for large and rapidly changing phase errors during the observations. We imaged the continuum delay calibrator using DIFMAP[34] in order to produce a source brightness model composed of CLEAN component. This was used in the AIPS task FRING to find the multi-band group delay and phase residuals.

The method consisted of transferring the multi-band delay solutions of the continuum delay calibrator and the phase rate solution of a strong $H_2O$ maser channel, which is copied from the K band to the calibration solutions of the SiO masers at high frequency band. The solutions were multiplied by the ratio between low and high frequencies. This method is the basis of non-integer SFPR[13]. The main point of the SFPR and non-integer SFPR method is transferring the low frequency phase solution to the high frequency[11–13]. Non-integer SFPR enables to make high frequency VLBI images for maser lines, even though a continuum delay calibrator is weak at the high frequency. We adopted a signal-to-noise threshold of maser spot identification to be 5–10, dependent on the image noise level[20], in the AIPS task SAD.

**Another astrometrically registered maps for comparison**. The monitoring results of February 27, 2016 ($\varphi = 0.63$) 1 month prior to March 27, 2016 ($\varphi = 0.67$) are presented in Supplementary Figs. 1, 2, 3 and Supplementary Table 2. The basic observation set-up is the same, and the system noise temperature at the K, Q, W, and D bands were up to 130, 200, 300, and 450 K. The result of the registered integrated velocity-intensity map at February 27, 2016 is shown in Supplementary Fig. 1. We could not obtain the 129.3 GHz SiO maser image because the fringe solution was not obtained at this epoch. The velocity–position maps for each frequency maser are represented in Supplementary Fig. 2. The flux of the single dish and VLBI were about 1.5 times stronger at 43.1 and 42.8 GHz, but the morphology on the spot distribution was similar to that of March 27, 2016 ($\varphi = 0.67$). The 86.2 GHz maser shows a stronger flux than that of 43.1 and 42.8 GHz masers at both two epochs.

The SiO maser ring fitting results are shown in Supplementary Table 2. The size distribution tendency of each frequency radii was the same within the error range as those of March 27, 2016 ($\varphi = 0.67$) (42.8 < 43.1 < 86.2 GHz ring radius). Supplementary Fig. 3 shows the registered map of the SiO masers and the spot distribution histograms from the center of the fitting in each maser line. The 42.8 GHz SiO maser is located in the innermost region, and the 86.2 GHz is at the outermost region. The stellar position difference determined from the ring fitting center using all SiO maser spots is ΔR.A. = 1.23 and ΔDec. = −0.23 mas between these two observations (Table 1 and Supplementary Table 2).

**Recovering fluxes**. Figure 2 and Supplementary Fig. 2 include the spectra of the VLBI total power and recovered flux (dotted and dashed lines) together with a single-dish (solid line) spectra obtained from the closest date to the VLBI observations. Because the beam size of the VLBI is very narrow compared to that of the single dish, there is a missing flux in the VLBI observations. Also, the different baseline removal method causes flux difference between the VLBI total power and single-dish spectra. In the case of a single-dish observation (position-switching mode), we observe the empty sky to remove the baseline from the target spectrum. On the other hand, the VLBI is calibrated using bandpass data obtained from fringe finder observations.

Moreover, the maser clumps with a larger angular size compared to the VLBI beam are resolved out in the cross-correlated spectra (recovered flux). The ratios of the VLBI total power to single-dish flux and recovered values (the ratio of recovered to total power fluxes) are listed in Supplementary Table 3. In the 42.8 GHz SiO maser, the relatively high recovery values seem to originate from a more compact maser spot in the 42.8 GHz maser than that in the 43.1 GHz maser.

**Astrometric uncertainty**. Our registered maser maps consist of the conventional PR maps of the $H_2O$ maser, with respect to J1833-2103, and SFPR maps of the SiO masers, with respect to the $H_2O$ maser positions determined from the PR maps. Therefore, we need to analyze astrometric errors related to these two kinds of registered maps. The propagation of the phase errors in the PR and SFPR results has been thoroughly studied using analytic and numerical methods[11,13,35,36]. Our observational parameters, such as 4 min for the source switching cycle time and 6.06° for the separation angle, yielded a somewhat large amount of errors in the PR maps. Following the formulae in previous work[35], we can estimate the residual phase errors per baseline in the PR method: ~90° phase noise from the dynamic tropospheric terms, ~120° phase noise from the static tropospheric terms, ~2° phase noise from the dynamic ionospheric terms, and ~15° phase noise from the static ionospheric terms. In addition, the large uncertainty of the antenna positions (cm order) of the KVN yields about 20° of phase noise in our PR results. The tropospheric contributions to the error budget can be removed in the SFPR results of the KVN due to the simultaneous multi-frequency observational capability[11]. Therefore, only the ionospheric terms contribute to the phase errors in the SFPR method, which yield 5° of phase noise from the dynamic ionospheric terms and 30° of phase noise from the static ionospheric terms.

Following the astrometric accuracy of the $H_2O$ maser positions in KVN PR maps, it can be estimated to be around 2 mas[13], which causes an uncertainty of 1 mas between the 22.2 GHz $H_2O$ and the 43.1 GHz SiO, 1.5 mas between the 22.2 GHz $H_2O$ and the 86.2 GHz SiO, and 1.7 mas between the 22.2 GHz $H_2O$ and 129.3 GHz SiO. The uncertainty between the SiO maser transitions can be also derived: 9 µas between the 42.8 GHz SiO and the 43.1 GHz SiO, 1 mas between the 43.1 GHz SiO and the 86.2 GHz SiO, and 1.3 mas between the 43.1 GHz SiO and the 129.3 GHz SiO maser. Obtaining precise absolute coordinates of the $H_2O$ maser emission using the PR at low frequency is important for the accurate estimation of the position of the central star assumed to be located at the center of the SiO maser distribution, which is determined by the SFPR method. However, the uncertainty of the absolute astrometry will be large due to the large separation between the target and delay calibrator. This could be improved by including geodetic blocks or using GPS data to reduce the residual zenith path-length error.

**Ring fitting of the SiO masers**. We fitted the SiO maser ring structure of 43.1, 42.8, 86.2, and 129.3 GHz maser lines using the least-squares fitting method adopting the IDL online procedure "mpfitellipse.pro" (http://cow.physics.wisc.edu/~craigm/idl/idl.html). We did not apply the weights function because we assumed that the SiO maser is a perfect circular structure. Smallest chi-square error was selected as the best fitting result. The ring fitting results were expressed as a single center position with the ring radius range. The results of the estimated ring fitting uncertainty are listed in Table 1 and Supplementary Table 2.

**Four single-band maps of the 129.3 GHz SiO maser**. We also detected the 129.3 GHz SiO maser at four epochs: March 14, 2015 ($\varphi = 0.11$), November 28, 2015 ($\varphi = 0.50$), December 30, 2015 ($\varphi = 0.54$), and December 30, 2016 ($\varphi = 1.10$), but the astrometric information was not confirmed at these epochs (Supplementary Fig. 4 and Supplementary Table 4). It is because we could not trace delay rate from non-integer SFPR according to the weather and observation conditions. The system temperatures were 620, 600, 550, and 500 K, the peak fluxes were 15.51, 7.27, 9.27, and 4.54 Jy beam$^{-1}$, and the rms values were 60.34, 25.83, 20.42, and 18.65 Jy beam$^{-1}$ m s$^{-1}$ at each epoch, respectively. Compared with the 129.3 GHz map of Fig. 2, the blue-shifted maser features in Region D are the strongest on $\varphi = 1.10$. The red-shifted features in South (B') appear in all phases and show the strongest intensities among the other features near the minimum phase ($\varphi = 0.50, 0.54$, and 0.67). However, the feature in Region C did not appear at three epochs in 2015. It seems that the feature in Region C is newly generated at $\varphi = 0.67$. Region B was the strongest at the $\varphi = 0.11$ phase and tends to decrease over time that looks independent of the pulsation cycle. In contrast, Region A is the strongest in the last phase at $\varphi = 1.10$. Region A becomes richer and stronger from $\varphi = 0.11$ to 1.10. Relatively, maser spots are distributed widely at $\varphi = 1.10$ compared to the other epochs, and its ring size is the largest.

Figure 4 shows the variations of the ring radii of each SiO maser line together with the optical light curve obtained from AAVSO. The ring radius of the 129.3 GHz maser shows the largest size around the optical maximum compared to those of the other epochs. On the other hand, the ring radii of the 43.1, 42.8, and 86.2 GHz masers do not show the variations according to the optical phases. It is probably due to the differences in the angular resolution of the KVN at different frequencies. Namely, the angular resolution of the KVN at 129.3 GHz (1.0 mas) can resolve the variations of the ring radii (over the error range of the ring fitting, etc.) compared to the relatively low resolutions of the KVN at 43.1, 42.8, and 86.2 GHz. It is difficult to see the variations of meaningful ring size at 43.1, 42.8, and 86.2 GHz with the phase within the fitting error range and the limited spatial resolution of the KVN. The SiO ring sizes of each epoch are shown in Table 1 and Supplementary Tables 2 and 4.

**Data availability**. Raw data were generated at the Korea–Japan Correction Center (KJCC) in Daejeon. Derived data supporting the findings of this study are available from the corresponding author on request.

## Acknowledgements
This work was supported by the Basic and Fusion Research Programs (2014–2017). We are grateful to all of the staff members at KVN who helped to operate the array and the single dish telescope and to correlate the data. The KVN is a facility operated by KASI (Korea Astronomy and Space Science Institute), which is under the protection of the National Research Council of Science and Technology (NST). The KVN operations are supported by KREONET (Korea Research Environment Open NETwork), which is managed and operated by KISTI (Korea Institute of Science and Technology Information). H.I. was supported by the KASI Commissioning Program. In this research, we have used information from the AAVSO International Database that operates at AAVSO Headquarters, 25 Birch Street, Cambridge, MA 02138, USA.

## Author contributions
S.-H.C. led the overall project. D.-H.Y. (dhysgr@gmail.com) reduced the data assisted by Y.Y. and R.D., and wrote the manuscript with revisions by S.-H.C. and H.I., Y.K.C., M.R., J.K., D.K., and H.Y. performed data verification. D.-Y.B. provided advice for the 4 band

VLBI system setting and observations. All authors discussed the data analyses and the results and commented on the manuscript.

## Additional information

**Supplementary Information** accompanies this paper at https://doi.org/10.1038/s41467-018-04767-8.

**Competing interests:** The authors declare no competing interests.

**Reprints and permission** information is available online at http://npg.nature.com/reprintsandpermissions/

**Publisher's note:** Springer Nature remains neutral with regard to jurisdictional claims in published maps and institutional affiliations.